\documentstyle[prl,aps,twocolumn,floats]{revtex}
\begin{document}
\twocolumn[\hsize\textwidth\columnwidth\hsize\csname
@twocolumnfalse\endcsname
\rightline{Preprint BROWN-HET-1022, CLNS 95/1368, MIT-CTP-2485 (1995)}
\begin{flushright}
{hep-ph/9511349, November 1995}
\end{flushright}
\vspace*{3mm}
\begin{center}
{\bf Dynamical Breaking of CPT Symmetry in Defect Networks and
Baryogenesis}
\end{center}
\bigskip
]
\centerline{\large\rm Tomislav Prokopec\footnote{\rm e-mail:
tomislav@hepth.cornell.edu}}
\begin{center}
{\it Newman Laboratory for Nuclear Studies, Cornell University,
Ithaca NY 14853, USA}
\end{center}
\centerline{\large\rm Robert Brandenberger\footnote{\rm e-mail:
rhb@het.brown.edu}}
\begin{center}
{\it Physics Department, Brown University, Providence RI 02912, USA}
\end{center}
\centerline{\large\rm Anne C. Davis\footnote{\rm e-mail:
A.C.Davis@damtp.cam.ac.uk}}
\begin{center}
{\it D.A.M.T.P., Silver Street, Cambridge University,
Cambridge CB3 9EW, UK}
\end{center}
\begin{center}
       {\large\rm and}
\end{center}
\centerline{\large\rm Mark Trodden\footnote{\rm e-mail:
trodden@ctpa04.mit.edu}}
\begin{center}
{\it Center for Theoretical Physics,  Laboratory for Nuclear Science and\\
Department of Physics, Massachusetts Institute of Technology,
Cambridge MA 02139, USA}
\end{center}

\date{\today}


\begin{center}
{\bf Abstract}
\end{center}

Based on a study of {\it charge,} (C), {\it parity} (P) and
{\it time reversal} (T) symmetries we show how a CP violating
network of defects in the early Universe may bias
baryon number production. A static network, even though it violates CP,
respects CPT and hence does not bias baryon number production.
On the other hand, the {\it ordering dynamics\/} of defects in a
network, governed by the interplay of string tension, friction,
inertia and the expansion of the Universe, results in the
dynamical breakdown of CPT symmetry and may lead to a net baryon
number production.

\vspace*{5mm}

\noindent {\bf 1. Introduction}

In Ref. \cite{BDPT95} (see also Refs. \cite{BDT94} and \cite{BD92}) we proposed
an
alternative electroweak scale baryogenesis mechanism which does not require the
electroweak phase transition to be first order, but which instead makes
essential use of the nontrivial dynamics of CP
violating cosmological defect networks.
The crucial third Sakharov condition for baryogenesis, the departure from
thermal equilibrium, is achieved by the out-of-equilibrium motion of the
defects
through the plasma in the expanding Universe. Some of the issues related to CP
and CPT violation were only touched on in
Ref. \cite{BDPT95}. The main goal of
this letter is to clarify these issues.

In Ref. \cite{BDPT95} we assumed that at the time of the electroweak phase
transition there is a network of defects (e.g. cosmic strings) which were
produced at an energy scale slightly higher than the electroweak scale and in
the cores of which the electroweak symmetry is unbroken (for some concrete
models see e.g. Ref. \cite{TDB95}). As the defects move through the primordial
plasma, a nonvanishing net baryon number can be generated. As in many of the
other electroweak baryogenesis mechanisms based on critical bubbles produced in
a first order phase transition generating a net baryon number (see e.g. Refs.
\cite{turokreview,CKNreview}), we assume that there is extra CP violation in
the Higgs sector (requiring us to consider extensions of the minimal standard
model). In analogy to how the CP violating phase changes when a bubble wall
passes over a point $p$ in space, generating a net baryon number density at
$p$, baryons will be generated when a wall of a topological defect!
!
 passes over $p$. More precisely,
in the case of local baryogenesis,
antibaryons will be generated when $p$ enters the defect (because the CP
violating phase changes in opposite direction to what happens when $p$ changes
from being in the false to being in the true vacuum), and an equal number of
baryons are generated when $p$ exits the defect. However, inside the defect the
antibaryons are converted to leptons {\it via\/} sphaleron processes (since the
latter are un-suppressed inside the defects as long as the defects are
sufficiently thick), and therefore the net result of the dynamics is to produce
a nonvanishing baryon number density. Obviously, both diffusion and the
expansion of the Universe (without which there would be no defect network) play
a crucial role in this mechanism. We have for simplicity described a local
baryogenesis scenario in which
CP violation and baryogenesis occur at the same spatial point. A similar
description holds for non-local baryogenesis scenarios.

It will be seen that our mechanism is closely related (with notable
differences which will become clear in the course of the letter) to the idea of
{\it spontaneous baryogenesis\/} of Ref. \cite{CohenKaplan} which asserts that
if there is a field -- named by the authors the {\it ilion\/} -- in the early
Universe which couples to the baryonic current, then either the expansion of
the Universe alone or a potential for this field could cause it to evolve in a
non-trivial manner, thus biasing baryon number production.

Dine {\it et al\/} \cite{DineHuetSingletonSusskind} and
Cohen, Kaplan and Nelson \cite{CKNspontaneous} realized that
a CP violating relative Higgs phase $\theta$
of a two Higgs doublet model may `play the role' of the ilion field. There are,
however, notable differences:
the phase $\theta$ couples {\it via\/} a derivative coupling
to an axial current $j^\mu_5$ so that the effective Lagrangian containing a
term $\propto (\partial_ \mu\theta ) j^\mu_5$ is CPT conserving.
Nevertheless, CPT is
violated {\it dynamically}.  As the Universe super-cools in the false symmetric
phase,  critical bubbles of `true' vacuum nucleate and grow driven by the
release of latent heat. This bubble growth is the mechanism
for dynamical violation of CPT symmetry and a net baryon number may be created.
(The CPT transformed situation would comprise collapsing bubbles, which is
clearly thermodynamically forbidden.)

In this letter we will show that the mechanism of Ref. \cite{BDPT95} is an
alternative way to obtain dynamical
CPT violation, based on the ordering dynamics of defect networks
and hence it is a realization of the original idea of spontaneous baryogenesis.
Our case, however, is slightly more complicated than the original scenario in
that the net change in $\theta$ vanishes.

\vspace*{5mm}

\noindent {\bf 2. Is there a CPT Paradox?}

Let us review the mechanism of Ref. \cite{BDPT95} in a bit more detail. The
first assumption is that cosmic strings are produced at a
phase transition above the electroweak scale, {\it i.e.\/} at an energy scale
$T_{CS}$ which satisfies
$T_{CS}>T_{ew}\simeq 100$GeV, but which is not too high so that by the
time of the electroweak phase transition the strings are not too
diluted. Some particle physics models in which this mechanism can be
implemented are discussed in Ref. \cite{TDB95}. Another candidate model is one
in which
the supersymmetric grand unified phase transition occurs (as a consequence of
the presence of
flat directions in the grand-unified Higgs potential) at around
$T\sim 1$TeV \cite{LythStewart}, \cite{BarrieroCopelandLythProkopec}.
We also need to assume that after the electroweak phase transition the
electroweak symmetry
$SU(2)_L\times U(1)_Y$ is preserved in the core of the strings,
{\it i.e.\/} the electroweak Higgs expectation value vanishes.
If there is extra CP violation in the theory, realized through {\it e.g.\/}
explicit CP violation in the Higgs sector of a two Higgs doublet model,  the
CP violating relative Higgs phase $\theta$ will change across the string in a
definite manner, just like in the case of bubble growth in a first order phase
transition \cite{TurokZadrozny}, \cite{CKNspontaneous},
\cite{ClineKainulainenVischer}. In
this case strings are not CP invariant field configurations and the CP
conjugate configurations are not solutions to the field equations.
They would thus have a much higher energy.
That is how the explicit CP violation in the Higgs
sector is manifest in a string network.
In this respect the `ground state' is CP violating.
Analogous investigations apply to
domain walls and monopoles. However, monopoles lead to a large
volume suppression factor for electroweak baryogenesis. Domain walls, although
they would be more effective than cosmic strings from a geometric point of view
(more volume in which baryogenesis can take place) suffer from the problem
of energy dominance: unless one invokes as a remedy {\it e.g.\/} additional
symmetry breaking that would destroy them, they would eventually dominate the
energy density of the Universe. Cosmic strings, on
the other hand, since they reach a scaling solution do not suffer from this
problem. Note that the arguments we will present here concerning the basic
baryogenesis mechanism are general in the sense
that they apply to any defect network.

The CP violating relative Higgs phase $\theta$ changes across
the core of the string (this is for example worked out for a spherical bubble
in Ref.
\cite{ClineKainulainenVischer}). If we set it to
{\it zero\/} outside the string it acquires a definite
sign on the wall and in the core of the string, say $\theta \geq 0$ everywhere.
The length over which $\theta$ varies,
which specifies
the thickness of the `wall' and core size $L$ of the defect, is given by the
electroweak
scale but is somewhat model
dependent. Provided $L$
is sufficiently large to accommodate the sphaleron of typical size
$(g_w^2T)^{-1}$, {\it i.e.}
$L > (g^2_wT)^{-1}$, it is plausible that the sphaleron
rate is un-suppressed in the core of the string \cite{Perkins}. In this case,
the rate of sphaleron transitions per
unit volume is given by $\Gamma_{sph}/V=\kappa_{sph}(\alpha_wT)^4$,
$\kappa_{sph}\sim 1$ \cite{AmbjornKrasnitz},
and the standard baryogenesis mechanism will apply
\cite{CKNtransport,JPTleptons,CKNdiffusion,JPTthin,JPTthick}.

A static string configuration is not CP invariant. The phase
$\theta$ has the following transformation properties:
it is even under parity (P), odd under charge conjugation (C)
and odd under time reversal (T), so that under CP
$\theta(x, t) \rightarrow  -\theta(-x, t)$ and under T $\theta(x, t)
\rightarrow - \theta(x, -t)$, and thus under CPT
$\theta(x, t) \rightarrow \theta(-x, -t).$ Hence
$\partial_\mu\theta$ is odd under CPT. (Note that $\partial_\mu\theta$
transforms as a vector field.)

How the string interacts with the plasma can be modeled by a
term in Lagrangian of the form
\begin{equation}
{\cal L}_\theta \propto (\partial_\mu\theta ) j^\mu_5
\label{eq:lagrangiantheta}
\end{equation}
where $j^\mu _5=\bar \Psi\gamma^\mu \gamma_5\Psi$ is the axial current.
This term can be obtained by a specific local rotation of fermions with
rotation angle proportional to $\theta$.
(For the exact form of
(\ref{eq:lagrangiantheta}) in a two Higgs doublet model see \cite{JPTthin}.)

The axial current transforms under CPT as
$j^\mu_5(x^\nu)\rightarrow -j^\mu_5(-x^\nu)$ so that the Lagrangian
(\ref{eq:lagrangiantheta}) is invariant under CPT as it should be. We conclude
that a static string under CPT transformation transforms into itself,
and hence can be considered to be its own `anti-particle'.

We will now relate this conclusion to the CPT theorem.
Recall that the CPT theorem states that any Lorentz invariant Lagrangian
${\cal L}(x)$ transforms under CPT into ${\cal L}^\dagger (-x)$ and hence, if
hermitean, ${\cal L}$ is invariant under CPT \cite{Lee}. A consequence of
this theorem is that any stable configuration must be either its own
`anti-particle' or have an `anti-particle' of exactly the same energy. Since
defects are not CP eigenstates,  the anti-defects have larger energy, hence
they must be their own `anti-particles'. Indeed, under CPT a static string
transforms into itself. This is in agreement with the conclusion we have
reached above. We have now established CPT invariance of a static defect
network and its consistency with the CPT theorem.

How is it then possible that one gets any baryogenesis from defects? If a
string moving in one direction produces a net number of baryons, then it seems
that based on the CPT theorem the same string moving in the opposite direction
should produce the same number of antibaryons. But our microphysical mechanism
of baryogenesis is independent of the direction of motion of the string.
In the rest of the paper we will argue that this apparent CPT paradox can be
resolved taking into account the dynamics of the string network.

\vspace*{5mm}

\noindent {\bf 3. Resolution}

For a static string there is no paradox: by the CPT theorem a static string as
its own `antiparticle' cannot
generate a net baryon number.
The same conclusion holds for a moving string in the absence of dissipation:
the baryon density induced at the trailing edge of the defect exactly
cancels the antibaryon density induced at the leading edge. As we will now
see a net baryon number density results only if dissipation is effective.

As argued in \cite{BDPT95}, a moving string
drives plasma out of equilibrium through coupling to the plasma {\it via\/}
a term of the form (\ref{eq:lagrangiantheta}).
Thermal equilibrium is restored through dissipative processes in the plasma, an
example being the biased sphaleron process
\begin{equation}
\dot n_B\propto -\Gamma_{sph}\mu_B,
\label{restauration of eq}
\end{equation}
where $\Gamma_{sph}$ is the
sphaleron rate, $n_B, \mu_B$ are baryon number density and the corresponding
chemical potential. These processes all violate {\it time
reversal\/} symmetry (T) and since they conserve CP, CPT is violated.
This dynamical CPT violation should not surprise us too much: both when there
is a surplus of particles
over anti-particles and {\it vice versa\/} (CPT conjugate case), the
out of equilibrium processes such as (\ref{restauration of eq})
will tend to restore a thermal equilibrium with
equal numbers of particles and antiparticles.
This out-of-equilibrium dissipative CPT
violation makes the crucial difference between static and moving strings:
{\it moving strings induce an effective CPT symmetry violation} analogous
to the effective CPT violation induced by the dynamics of the {\it ilion}
field of Ref. \cite{CohenKaplan}.

We will now describe a non-local baryogenesis scenario which,
when compared with local baryogenesis scenarios, typically
dominates baryon production.
For {\it thick-walled\/} defects, for which particles scatter typically many
times as they move across the phase boundary (wall), due to imperfect
transport and finite decay time, the  field  $\partial_\mu\theta$
(sometimes called `charged potential')
will not be perfectly screened. In the case of {\it thin-walled\/}
defects for which the scattering length exceeds the phase boundary
thickness, a coherent quantum mechanical reflection will take place and
source axial current in  the vicinity of the defect; transport and decays
will determine the destiny of this current, {\it e.g.} how it thermalises.
One can show that under rather weak conditions
(essentially sub-sonic velocity of the string is the necessary and sufficient
condition), for both thin- and thick-walled defects, a diffusion tail of
particle {\it minus\/} anti-particle excess forms in front of both
the trailing and the front edges of a moving defect. The excess that
overlaps the string core biases baryogenesis since in the core the
sphaleron rate is un-suppressed, while the excess in front of the defect
has no effect. We will now consider some of the aspects of
this model related to symmetry conservation/violation.

Consider a segment of a string moving through the plasma which
was set into motion by some initial kick. According to
\cite{BDPT95}, the string leaves a trail of baryons
in its wake. But also it slows down due to the friction force that plasma
exerts and eventually stops. Hence, baryon production also stops. This process
can be viewed as follows: a force that put the string in motion must be of
non-equilibrium origin; the string then approaches equilibrium as it slows
down. There is no contradiction since baryons are produced out of equilibrium.
The key question now becomes: {\it What is the force that constantly
kicks the strings in the early Universe?}

\vspace*{5mm}
\noindent {\bf 4. Out of Thermal Equilibrium}

Strings are formed at a phase transition above the electroweak scale. The
strings are a measure of the deviation of the field configuration from being in
perfect thermal equilibrium. In the hypothetical limit of infinite transition
time, no strings would remain. Immediately after the transition,
the ordering dynamics is governed mostly by the string tension, inertia and
friction; the expansion of the Universe is irrelevant. This initial stage is
called friction-dominated regime.
As the Universe expands, the friction decreases (since the plasma density is
redshifted).
Also, the long strings that typically traverse many horizons are stretched.
Eventually, the expansion of the
Universe becomes a more important `damping force' for network dynamics than the
friction, and the network enters the so called scaling regime in which the mean
separation of strings remains proportional to the Hubble radius \cite{strings}.

In both regimes strings are evolving out of equilibrium. Moving strings drive
the surrounding plasma out of
equilibrium. This ordering dynamics of string straightening never stops in
an expanding Universe simply because the correlation length keeps growing.
Correlations in the phases of the scalar field are established through
dissipative processes like string intercommutation and string loop decay into
gravitational radiation. The expansion of the Universe is hence crucial
because {\it it keeps the network out of thermal equilibrium when
it reaches a scaling regime}. In addition
it cools down the Universe and hence provides the arrow of time.

Another way to demonstrate the out-of-equilibrium nature of the defect
network dynamics is to consider what happens to a distribution of
strings in the scaling regime if we suddenly let the Universe contract.
On scales smaller than the Hubble radius at the time when the contraction
starts, the phase coherence will be maintained. Hence, the final string
configuration at the end of the contraction will be different from the
initial configuration at the beginning of the expansion (assuming that
the scale factor of the Universe at these two times is the same).

What about thermal fluctuations? Thermal excitations may generate string
loops (that decay quickly). Since these loops are equilibrium configurations,
there is no meaningful definition of time arrow and hence no dissipation and
time reversal violation which are necessary for net baryon production.
Indeed these thermal loops cannot excite net axial current and therefore
no baryogenesis is possible.

\vspace*{5mm}
\noindent {\bf 5. Conclusions}

We have argued that any network of defects dynamically breaks CPT symmetry.
When these defects couple to the left handed fermion current or in fact
any current that is not orthogonal to it (an example is the axial current in
(\ref{eq:lagrangiantheta}) which can be decomposed into the left-handed and
right-handed fermion currents) in a CP violating manner ({\it via\/} a CP
violating
field as in (\ref{eq:lagrangiantheta})), and when in motion, they may
bias baryon number production {\it via\/} the sphaleron processes in the core
of defects.

We now compare this mechanism with the most popular model of electroweak
baryogenesis in which baryon production occurs at a first order electroweak
phase transition: on or around the phase boundary
of a growing bubble axial currents are induced, driving fermions in
the plasma out of equilibrium; and the bubbles expand due to release of latent
heat. The motion of cosmic strings, on the other hand, is generated by the
string tension and by the expansion of the Universe, both of which
tend to straighten them. The dynamics is intrinsically out-of-equilibrium and
does not approach an equilibrium configuration, at least on scales larger or
equal to the string correlation length.

In conclusion,
{\it when at a phase transition CP violating defects are produced,
{\rm ordering dynamics\/} drives the system out of equilibrium locally,
leading to dynamical CPT violation. In
conjunction with CP violation this biases baryon number production.}

\vspace*{3mm}
\centerline{\bf Acknowledgments}
The research of RB is funded in part by DOE Grant DE-FG02-91ER40688-Task A,
that of AD by PPARC,
TP acknowledges funding from PPARC and NSF, and
MT is supported in part by funds provided by the US Department of Energy
(D.O.E.) under cooperative research agreement \#DF-FC02-94ER40818.
We thank Michael Joyce, David Lyth and Neil Turok for useful discussions.

\end{document}